\documentclass[a4paper,10pt]{article}

\usepackage{graphicx}
 \usepackage{amsmath}
 \usepackage{amsfonts}
 \usepackage{amssymb}

\newcommand{\dr}[2]{\frac {{\rm d} {#1}} {{\rm d} {#2}}}

\newcommand{\dril}[2]{{{\rm d} {#1}} / {{\rm d} {#2}}}

\begin{document}

\setlength{\oddsidemargin} {1cm}
\setlength{\textwidth}{18cm}
\setlength{\textheight}{23cm}

\title{A (giant) void is not mandatory to explain away dark energy with a
Lema\^itre--Tolman model}

\maketitle

\author{{\bf Marie-No\"elle C\'el\'erier$^{1,a}$, Krzysztof Bolejko $^{2,3,b}$ and
Andrzej Krasi\'nski$^{3,c}$} \\
{\small $^1$ Laboratoire Univers et TH\'eories (LUTH), Observatoire de Paris,
CNRS, Universit\'e Paris-Diderot, 5 place Jules Janssen, 92190 Meudon, France} \\
{\small $^2$ Department of Mathematics and Applied Mathematics, University of Cape Town, Rondebosch 7701, South Africa} \\
{\small $^3$ Nicolaus Copernicus Astronomical Centre, Polish Academy of Sciences, Bartycka 18, 00 716 Warszawa, Poland} \\
{\small e-mail: $^a$ marie-noelle.celerier@obspm.fr} \\
{\small $^b$ bolejko@camk.edu.pl} \\
{\small $^3$ akr@camk.edu.pl \\}}


\begin{abstract}

Lema\^itre -- Tolman (L--T) toy models with a central observer have been used
to study the effect of large scale inhomogeneities on the SN Ia dimming. Claims
that a giant void is mandatory to explain away dark energy in this framework are
currently dominating. Our aim is to show that L--T models exist that reproduce a few features of the $\Lambda$CDM model, but do not contain the giant cosmic void.
We propose to use two sets of data -- the angular diameter distance together
with the redshift-space mass-density and the angular diameter distance together
with the expansion rate -- both defined on the past null cone as functions of
the redshift. We assume that these functions are of the same form as in the
$\Lambda$CDM model. Using the Mustapha--Hellaby--Ellis algorithm, we numerically
transform these initial data into the usual two L--T arbitrary functions and
solve the evolution equation to calculate the mass distribution in spacetime.
For both models, we find that the current density profile does not exhibit a
giant void, but rather a giant hump. However, this hump is not directly
observable, since it is in a spacelike relation to a present observer.
The alleged existence of the giant void was a consequence of the L--T models
used earlier because their generality was limited a priori by needless
simplifying assumptions, like, for example, the bang-time function being
constant. Instead, one can feed any mass distribution or expansion rate history
on the past light cone as initial data to the L--T evolution equation. When a
fully general L--T metric is used, the giant void is not implied.

\end{abstract}

\section{The historical background of the problem} \label{int}

In the framework of the homogeneous and isotropic standard cosmological model,
the dimming of the type Ia supernovae as compared to their expected luminosity
in an Einstein--de Sitter model is interpreted as a consequence of an assumed
accelerated expansion of the Universe. This leads to the widespread belief in a
`dark energy' component currently dominating the energy budget of our Universe.
But this is not the only possible explanation of the SN Ia observations.

Shortly after the discovery by Riess {\em et al.} (1998) and Perlmutter {\em et
al.} (1999) of the supernova dimming, it was proposed by several authors that
this effect could be due to the large-scale inhomogeneities (Pascual-S\'anchez
1999; C\'el\'erier 2000; Tomita 2000; 2001a \& b). After a period of relative
disaffection, this proposal experienced a renewed interest about five years ago.

Three methods have been used to implement such a proposal: computation of
backreaction terms in the dynamical equations using an averaging procedure
proposed by Buchert (2000, 2001), calculations in the framework of a
perturbative scheme and the use of exact inhomogeneous models, in particular
those of Lema\^{\i}tre (1933) -- Tolman (1934) (L--T) (see C\'el\'erier 2007,
for a review).

The L--T model became rapidly popular for the purpose of mimicking `dark energy'
because it exhibits three interesting features: i) it is one of the few exact
solutions of General Relativity able to represent a physically consistent model
of the matter dominated era of the Universe, ii) among these few, it is the most
easily tractable from a computational point of view,
iii) it is not an alternative to, but a {\em generalisation of} the Friedmann
dust models (which are contained in it as a subcase), so can reproduce all the
Friedmann-based results, including those of the ``concordance'' $\Lambda$CDM
model, with an arbitrary precision. For more on this, in relation to the main
subject of this paper, see the last section.

Three classes of models have been constructed with the L--T solution: i) models
where the observer is located at the centre of a single L--T universe (e.g.,
Iguchi {\em et al.} 2002; Alnes {\em et al.} 2006; Apostolopoulos {\em et al.}
2006; Bolejko 2008; Garcia-Bellido \& Haugb\o lle 2008a, 2009), ii) models where
the observer is located off the centre of such a universe (e.g., Schneider \&
C\'el\'erier 1999; Apostolopoulos {\em et al.} 2006; Alnes \& Amarzguioui 2007)
\footnote{However, these authors have shown that the CMB data put very stringent
limits on the distance of the observer from the centre of the model or/and on
the amplitude of the homogeneities in an off-centre observer model.} iii)
Swiss-cheese models where the holes are L--T bubbles carved out of a
Friedmannian homogeneous background (e.g., Brouzakis {\em et al.} 2007, 2008;
Biswas \& Notari 2008; Marra {\em et al.} 2007).

As will be recalled in Sec.~\ref{lt}, an L--T model is defined by two
independent arbitrary functions of the radial coordinate, which can be fitted to
the observational data. However, in most of the models currently available in
the literature, the authors have artificially limited the generality by giving
the L--T initial-data functions a handpicked algebraic form (depending on the
authors' feelings about which kind of model would best represent our Universe),
with only a few {\it constant} parameters being left arbitrary -- to be adapted
to the observations.

 Another way in which the generality of the L--T models was artificially
limited was the assumption that the age of the Universe is everywhere the same,
i.e. that the L--T bang-time function $t_B$ is constant. With $t_B$ being
constant, the only single-patch L--T model that fits observations is one with a
giant void (Iguchi {\em et al.} 2002; Yoo {\em et al.} 2008). Conceptually there
is nothing wrong with a non-simultaneous big bang (even though this is a radical
qualitative difference with the FLRW models), but one should exercise caution
when referring to $(t -t_B)$ as the actual age of the Universe. The L--T model
is too simple to extend it up to instants earlier than decoupling. Therefore
$t_B$ should merely be regarded as a function that describes a degree of
inhomogeneity of the initial conditions rather than as the actual instant of
birth of the Universe.

The argument brought in defense of the constant $t_B$ assumption is this: a
non-constant $t_B$ generates decreasing modes of perturbation of the metric
(Silk, 1977, Pleba\'nski and Krasi\'nski, 2006), so any substantial
inhomogeneity at the present time stemming from $t_{B,r} \neq 0$ would imply
`huge' perturbations of homogeneity at the last scattering. This, in turn, would
contradict the CMB observations and the implications of inflationary models (we
deliberately do not give references here, to avoid blaming any single individual
for what seems to be a piece of conventional wisdom). However, these are only
expectations that should not be treated as objective truth until they are
verified by calculations. Such calculations have already been done, and it
turned out that the inhomogeneities in $t_B$ needed to explain the formation of
galaxy clusters and voids are of the order of a few hundred years (Bolejko {\it
et al.} 2005; Krasi\'nski and Hellaby 2004; Bolejko 2009). Then, on the basis
of Bolejko's (2009) models 4 and 5, one can calculate that for a structure of
present radius 30 Mpc this age difference between the oldest and youngest region
would generate CMB temperature fluctuations equal to $\Delta T/T = 3.44 \times
10^{-6}$ and $\Delta T / T = -2.35 \times 10^{-6}$, respectively. This is
well-hidden in the observational errors at the present level of precision. (In
the future, when, presumably, the precision will improve, these results may
possibly be used to {\em measure} the gradient of $t_B$.) So there is no
observational justification to the assumption $t_B =$ constant.

These are the reasons why, in the recent years, we have seen the increase in
popularity of void models, where the observer is
located at or near the centre of a large, huge, giant L--T void of size of up to
a few Gpc. Many authors have constructed classes of L--T models with a central local void
and have shown that they were able to fit the SN Ia and other cosmological data
provided the void is large (e.g., it has a diameter of 400 $h^{-1}$ Mpc
in Alexander {\em et al.} 2007) or even huge (e.g., 1.35 Gpc in Alnes {\em et
al.} 2006, $>$ 2 Gpc in Garcia-Bellido \& Haugb\o lle 2008a and 2009), depending on
the features of the particular model they had chosen. This contributed to the
spreading of the belief in the necessity of a `giant local void' to resolve the
`cosmological constant problem' with L--T models. However, as shown by Mustapha {\em et al.} (1998) and used as an illustration for the application to the supernova data and the `cosmological constant problem' by C\'el\'erier (2000), a given set of isotropic data can constrain only one of the two free functions of an L--T model and therefore, after fitting
the supernova data with a given L--T solution, we are left with plenty of room
to accommodate more observations.

Actually, a few authors discarded the central void hypothesis and proposed models with no
such void (e. g., Iguchi {\em et al.} 2002). Enqvist and
Mattsson (2007) even showed that the fitting of the SN Ia data can be better
with L--T models where the density distribution is constant on a constant-time
hypersurface than with the $\Lambda$CDM model (see also Bolejko 2008 and Bolejko
\& Wyithe 2009). Even though such a density distribution is not what is actually
observed at very large scales by astronomers (it is not even observable, being
in a spacelike relation to the central observer), the $\rho(t_0,r) =$ const
configuration vividly illustrates how misleading the FLRW-based geometrical
intuitions can be. A spatial distribution of matter can radically change with
time in consequence of an inhomogeneous expansion distribution in space. Our
models will provide more examples of this phenomenon, and we will come back to
this point in the conclusion.

Our aim here is to show, using two explicit examples reproducing the
observational features of the $\Lambda$CDM model, that
a giant void is not at all a necessary implication
of using L--T models. We propose to use input functions that can be derived from
observations.\footnote{One should be aware that there is a great deal of
phenomenology involved in interpreting the observations. For example, with
supernovae, the cosmological model predicts $D_L$, while what is actually
observed is the flux. We can deduce the absolute luminosity only on the basis of
some empirical methods. Similarly with galaxy number counts -- the cosmological
model predicts $m(z) n(z)$ where $m(z)$ is an average mass per source and $n(z)$
is number counts. The whole information about the galaxy evolution and their
mergers is encoded in $m(z)$ -- however in galaxy redshift surveys we observe
only $n(z)$. In this paper we do not focus on the problem of observations and
assume $D_L$ and $mn$ as in the $\Lambda$CDM.}. Our L--T toy models will be
constrained by the angular diameter distance together with the  redshift-space
mass-density or the angular diameter distance together with the expansion rate.

\begin{figure}
\begin{center}
\includegraphics[scale=0.65]{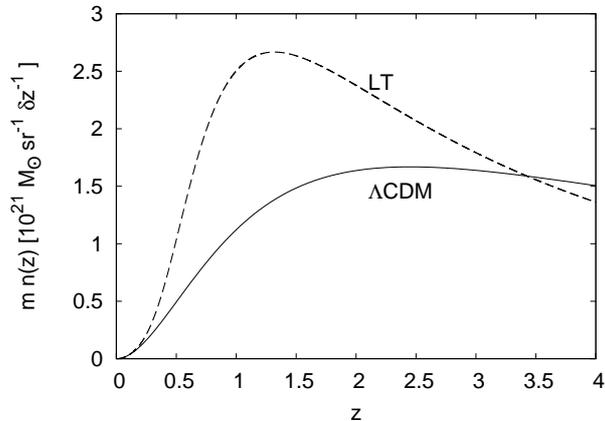}
\caption{The redshift-space mass-density in the $\Lambda$CDM model and in a
giant void L--T model (with radius of 2.96 Gpc and density contrast of 4.05).
At the redshift around 1 one should observe twice as many objects as in the
$\Lambda$CDM model.}
 \label{gnc}
\end{center}
\end{figure}

It should be noted that these functions have not the same form in the
$\Lambda$CDM model and in giant-void L--T models. For example, let us consider
the giant void model from Bolejko \& Wyithe (2009) with radius of 2.96 Gpc and
density contrast of 4.05. The redshift-space mass-density for this model and for
the $\Lambda$CDM model are shown in Fig.~\ref{gnc}. As seen, at $z \approx 1$
the difference between these two models is more than a factor of 2. Also, the
expansion rate as a function of redshift behaves differently (for details and
constraints coming from $H(z)$ see Bolejko \& Wyithe, 2009). Thus giant void
models have difficulties to mimic all the observational features of the dark
energy model (Zibin {\em et al.} 2008; Clifton {\em et al.} 2009).

In this paper, we show that {\it if} the observational data are properly fitted
to these $\Lambda$CDM functions, then a giant void is not mandatory to explain
them. In fact the L--T models that mimic our choice of observational features of
the $\Lambda$CDM model have a central Gpc-scale overdensity rather than an
underdensity. We emphasise that what we reproduce in our L--T model are not the
actual observational relations, but the parameters of the $\Lambda$CDM model
fitted to the observations -- which is not the same thing.

Note that this model is not designed to reproduce all the available cosmological
data, nor is it to be considered as the final model of our Universe. Its purpose
is to exemplify the proper use of L--T models and to show what can come out of
it. Moreover, it should be understood as tentative beyond the redshift range in
which the $\Lambda$CDM functions we use are robustly established.

As is usual in the study of L-T models, we choose to use a comoving and
synchronous coordinate system for the majority of this work. Such a coordinate
system is uniquely defined by the flow lines of the fluid and allows the
line-element to be written in a simple form. However, it is of course the case
that quantities such as energy density profiles on space-like volumes are
sensitive to the choice of hypersurface on which they are recorded. To
illustrate this dependence, and the effect of considering other
foliations, we also present our final results on a set of hypersurfaces in which
each fluid element is the same distance from the initial singularity along the world-lines of the dust particles. Such a choice allows us to consider the energy density of different regions when they are at the same age, and becomes a comoving and synchronous
coordinate system in the constant bang time models where giant voids have often been inferred.

In Sec.~\ref{lt} we recall briefly the main properties of the L--T solution and
give the equations to be integrated on the light cone. In Secs.~\ref{Dnmodel}
and \ref{DHmodel} we describe both models and give the results of our numerical
calculations. Section~\ref{disconc} is devoted to a discussion and a summary
 of our conclusions.

\section{The Lema\^itre -- Tolman solution} \label{lt}

The L--T model is one of two classes of spherically symmetric
solutions\footnote{For the presentation of the other class see Pleba\'nski and
Krasi\'nski (2006). It is called there the Datt -- Ruban solution (Ruban 1968,
1969). It has interesting geometrical and physical properties, but so far has
found no astrophysical application.} of Einstein's equations where the
gravitational source is dust. In comoving and synchronous coordinates, its line
element reads
\begin{equation}\label{2.1}
{\rm d} s^2 = c^2 {\rm d} t^2 - \frac {{R,_r}^2}{1 + 2E(r)}{\rm d} r^2 -
R^2(t,r)({\rm d}\vartheta^2 + \sin^2\vartheta \, {\rm d}\varphi^2),
 \end{equation}
where $E(r)$ is an arbitrary function and $R,_r = \partial R/ \partial r$. With
a vanishing cosmological constant, which is the case we consider here, $R(t,r)$
obeys the following first integral of one of the Einstein equations:
 \begin{equation}\label{2.2}
\frac{1}{c^2} {R,_t}^2 = 2E + \frac{2M}{R},
 \end{equation}
where ${R,_t} = \partial R / \partial t$ and $M = M(r)$ is another arbitrary
function of integration. $M(r)$ is the gravitational mass contained within the
comoving spherical shell at any given $r$, while $E(r)$ is the energy per unit
mass of the particles on that shell. Moreover, $E$ determines the space
curvature at each $r$-value. The mass density in energy units follows from the
other nontrivial Einstein equation and is
 \begin{equation}   \label{2.3}
\kappa \rho c^2 = \frac {2{M,_r}}{R^2R,_r}, \qquad {\rm where}\ \kappa = \frac
{8\pi G} {c^4}.
 \end{equation}

The solutions of (\ref{2.2}) can be written as
\begin{equation} \label{2.5}
R(t,r) = \frac{M(r)}{\chi(r)} \, \phi(t,r), \quad c \left( t - t_B(r) \right) =
\frac{M(r)}{(\chi(r))^{3/2}} \, \xi(t,r),
\end{equation}
where
\begin{enumerate}
\item when $E < 0$ (elliptic evolution):
\begin{eqnarray} \label{2.6}
   \chi(r) &=& -2 E(r),\\ \nonumber
   \phi &=& 1 - \cos \eta,\\ \nonumber
   \xi &=& \eta - \sin\eta,
 \end{eqnarray}

\item when $E = 0$ (parabolic evolution):
\begin{eqnarray}   \label{2.7}
   \chi(r) &=& 1,\\ \nonumber
   \phi &=& \eta^2/2,\\ \nonumber
   \xi &=& \eta^3/6,
\end{eqnarray}

\item and when $E > 0$ (hyperbolic evolution):
\begin{eqnarray} \label{2.8}
   \chi(r) &=& 2 E(r),\\ \nonumber
   \phi &=& \cosh\eta - 1, \\ \nonumber
   \xi &=& \sinh \eta - \eta,
 \end{eqnarray}
\end{enumerate}
where $\eta$ is a parameter (dependent on $t$ and $r$). The $t_B(r)$ is another
arbitrary function that appears as an integration `constant' and is interpreted
as the bang time, i.e., the Big Bang is not simultaneous for all values of the
$r$ coordinate.

Since all the formulae given so far are covariant under coordinate
transformations of the form $\widetilde{r} = g(r)$, one of the functions $E(r)$,
$M(r)$ and $t_B(r)$ can be fixed at will by the choice of $g$. Therefore, once
this choice is done, a given L--T model is fully determined by
two of these arbitrary functions.

However, as shown by Mustapha {\em et al.} (1998), and used as an illustration
for application to the supernova data and the `cosmological constant problem' by
C\'el\'erier (2000), a set of isotropic data corresponding to a given observable
can constrain only one of the two free functions, and therefore the fitting of
the supernova data, i.e. of the function $D_L(z)$, with a given L--T solution,
still leaves the other function free -- and available for fitting to another set
of data. Thus, we must also assume another set of initial conditions, e.g. the
redshift-space mass-density $m(z) n(z)$ or the expansion rate $H(z)$. We will
take these functions to be identical to the corresponding functions in the
$\Lambda$CDM model, assuming they reflect the observational data. By this, we
want to show that there is no antagonism between the inhomogeneous cosmology and
the $\Lambda$CDM model, and that the first can predict the same results as the
second even if $\Lambda=0$. However, whenever possible we present the real data
to show that there is still much room within observational errors for different
profiles.

Using the reciprocity theorem (Etherington 1933, Ellis 1971) the luminosity
distance can be converted to the angular diameter distance
\begin{equation}
D_A = R = \frac{D_L}{(1+z)^2}.
\end{equation}

For a ray issued from a radiating source and proceeding towards the
central observer on a radial null geodesic the following equation holds
\begin{equation} \label{2.10}
\frac{c \, {\rm d}t(r)}{{\rm d}r} = - \, \frac{R,_r(t(r),r)}{\sqrt{1 + 2E(r)}},
\end{equation}
and the equation for the redshift reads (Bondi 1947; Pleba\'nski \& Krasi\'nski
2006):

\begin{equation}
\frac{1}{1+z(r)} \frac{{\rm d}z(r)}{{\rm d}r} = \frac{1}{c}
\frac{R,_{tr}(t(r),r)}{\sqrt{1 + 2E(r)}}. \label{2.11}
\end{equation}
For later reference we will need also the following equation, which follows
easily from (\ref{2.10}) and (\ref{2.11}):
\begin{equation}
\frac{1}{1+z(t)} \frac{{\rm d}z(t)}{{\rm d}t} = - \frac{R,_{tr}(t, r(t))}
{R,_r(t,r(t))} \label{2.12}.
\end{equation}

\section{The L--T model with no central void} \label{mod}

Our aim is now to show we can design an L--T model with no central void able to
reproduce the angular diameter distance--redshift relation as inferred from the
SN Ia data, smoothed out as in the framework of a $\Lambda$CDM model. For this
purpose, we propose to use the following additional conditions to specify the
arbitrary functions $M(r)$, $E(r)$ and $t_B(r)$.

\subsection{The model defined by $D_A(z)$ and $m(z) n(z)$}\label{Dnmodel}

\subsubsection{The MHE procedure}\label{mhe-alg1}

The algorithm used to find the L--T model reproducing the $D_A(z)$ and $n(z)$
data was first developed by Mustapha, Hellaby \& Ellis (1997). Let us recall its
major steps and equations.

The radial coordinate $r$ is chosen so that, on the past light cone of $(t, r) =
(t_0, 0)$,
\begin{equation}
\frac{\widehat{R,_r}}{\sqrt{1+2E}} = 1. \label{corchoice}
\end{equation}
(Note: this choice of $r$ is possible only on a single light cone. In the
following, we always refer to the light cone of  $(t, r) = (t_0, 0)$.)
This choice of coordinates simplifies the null geodesic equation:
\begin{equation}
c \widehat{t}(r) = c t_0 - r,
\label{ngs}
 \end{equation}
where we denote quantities on this null cone by a hat. Furthermore, (\ref{2.3})
now becomes
\begin{equation}
(\kappa c^2/2) \widehat{\rho} \widehat{R}^2 = \frac{M,_r}{\sqrt{1 + 2E}}.
 \label{nlc_dens}
\end{equation}

The total derivative of the areal radius $R$ gives
 \begin{equation}
   \frac{{\rm d} \widehat{R}}{{\rm d} r} = \widehat{R,_r} + \widehat{R,_t} \,
     \frac{{\rm d} \widehat{t}}{{\rm d} r}.    \label{dRdr}
 \end{equation}
Using (\ref{2.2}), (\ref{corchoice}) and (\ref{ngs}), the above equation can be
written as
 \begin{equation}
\frac{{\rm d} \widehat{R}}{{\rm d} r} - \sqrt{1 + 2E} \, = - \frac{1}{c}
\widehat{R,_t} = \mp \sqrt{\frac{2M}{\widehat{R}} + 2E} \, .    \label{eq:Rhat2}
 \end{equation}

This can be solved for $E(r)$:
\begin{equation}
   1 + 2E = \left.\left\{ \frac{1}{2} \left[ \left( \dr {\widehat{R}} r
     \right)^2 + 1 \right] - \frac{M}{\widehat{R}} \right\}^2 \,\, \right/ \,\,
     \left( \dr{\widehat{R}} r\right)^2 \, .    \label{mheE}
 \end{equation}
Using (\ref{nlc_dens}) the above becomes
\begin{equation}
   \dr M r + \left({\displaystyle{\frac{\kappa c^2 \widehat{\rho} \widehat{R}}
   {\displaystyle{2 \, \dril {\widehat{R}} r}}}} \right) \, M =
   \left( {\displaystyle{ \frac{\kappa c^2 \widehat{\rho} \widehat{R}^2}
   {\displaystyle{4 \, \dril {\widehat{R}} r}}}} \right) \left[ \left(
   \dr {\widehat{R}} r\right)^2 + 1 \right] \, .     \label{mheM}
 \end{equation}

Matter density can be expressed in terms of $n(z)$ -- the observed number
density of sources in
the redshift space per steradian per unit redshift
interval. Thus, the number of sources observed in a given redshift interval and
solid angle is $n \, {\rm d}\Omega \, {\rm d}z$ and the total rest mass between
$z$ and $z + {\rm d}z$ is
 \begin{equation}
   \mathcal{M} = 4 \pi \, \widehat{m} \widehat{n} \, {\rm d} z,    \label{mndz}
 \end{equation}
where $\widehat{m}(z)$ is the average mass per source. On the other hand the
total rest mass between $r$ and $r + {\rm d} r$ is
 \begin{equation}
\mathcal{M} =  \widehat{\rho} \widehat{{\rm d}^3V} = \widehat{\rho} \frac{4 \pi
\widehat{R}^2 \widehat{R,_r}}{\sqrt{1 + 2E}} {\rm d} r,     \label{rhodV}
 \end{equation}
where $\widehat{{\rm d}^3V}$ is the proper volume on a constant time slice,
evaluated on the null cone.  Hence by (\ref{mndz}), (\ref{rhodV}) and
(\ref{corchoice})
 \begin{equation}
\widehat{R}^2 \widehat{\rho} = \widehat{m} \widehat{n} \dr z r\,.
\label{eq:rho_mn}
 \end{equation}
Finally, to find $r(z)$, the r.h.s. of  (\ref{2.11}) must be expressed in terms
of $\widehat{R}$ and $\widehat{n}(z)$. Differentiating (\ref{2.2}) with respect
to $r$ and substituting the result in the r.h.s. of (\ref{2.11}), we obtain
 \begin{equation}
   {\widehat{\frac{R,_{tr}}{\sqrt{1 + 2E} \,}}} =
   \frac{c^2}{\widehat{R,_t}} \left[ \frac{M,_r}{\widehat{R} \sqrt{1 + 2E} \,} -
   \frac{M}{\widehat{R}^2} + ({\sqrt{1 + 2E} \,}),_r \right].
 \end{equation}
The derivative of $\sqrt{1 +2E}$ follows from (\ref{mheE}). Then, replacing
$M,_r$ by (\ref{nlc_dens}) and using (\ref{eq:Rhat2}), the above equation can be
written as
 \begin{equation}
  \frac{1}{c} {\widehat{\frac{R,_{tr}}{\sqrt{1 + 2E} \,}}}
= - \left( \frac{1}{2}\kappa c^2  \widehat{\rho} \widehat{R} + \dr {^2
\widehat{R}} {r^2}\right) \, \left/ \, \left( \dr {\widehat{R}} r\right).\right.
\label{nullhz}
  \end{equation}
Now, from (\ref{2.11}):
 \begin{equation}
\dr {\widehat{R}} r \dr z r + \dr {^2\widehat{R}} {r^2} \, (1+z)
   =- \frac{1}{2}\kappa c^2 \widehat{\rho} \widehat{R} (1+z).
  \end{equation}
 Applying
 \[
   \dr {\widehat{R}} r = \dr {\widehat{R}} z \dr z r  \, , ~~~~~~
   \dr {^2\widehat{R}} {r^2} = \dr {\widehat{R}} z \dr {^2z} {r^2} +
   \dr {^2\widehat{R}} {z^2} \left(\dr z r\right)^2,
 \]
and integrating with respect to $r$, yields
 \begin{eqnarray}
   \int_{0}^{z} \dr {} r && \left[\dr {\overline{z}} r
          \dr {\widehat{R}} {\overline{z}} (1 + \overline{z}) \right]
 {\rm d} r \nonumber \\
&&
     = -\int_{0}^{z} \frac{1}{2}\kappa c^2 \widehat{\rho}(\overline{z})
       \widehat{R}(\overline z) (1 + \overline{z}) \dr r {\overline{z}}
       {\rm d} \overline{z}.
\end{eqnarray}
Using the origin conditions $[(\dril z r) (\dril {\widehat{R}} z)]_0 = [\dril
{\widehat{R}} r)]_0 = 1$, $z(0) = 0$ and (\ref{eq:rho_mn}) the above can be
rearranged to obtain
 \begin{eqnarray}
   \dr z r = && \left[ \dr {\widehat{R}} z (1 + z) \right]^{-1}
\nonumber \\
&&
 \times  \left\{ 1 - \frac{1}{2}\kappa c^2 \int_{0}^{z}
 \frac{\widehat{m}(\overline{z})
   n(\overline{z})} {\widehat{R}(\overline z)} \, (1 + \overline{z}) \,
   {\rm d} \overline{z} \right\}.   \label{eq:fstint}
  \end{eqnarray}
Finally:
 \begin{eqnarray}
   r(z)& = & \int_0^z \left[ \dr {\widehat{R}}{\tilde{z}} (1 + \tilde{z})
   \right]
   \nonumber \\
   &\times&
\left\{ 1 - \frac{1}{2} \kappa c^2 \int_{0}^{\tilde{z}}
\frac{\widehat{m}(\overline{z})
   n(\overline{z})} {\widehat{R}(\overline z)} \, (1 + \overline{z}) \,
   {\rm d} \overline{z} \right\}^{-1} \, {\rm d}\tilde{z} \,\, .
   \label{eq:rz_int}
 \end{eqnarray}

\subsubsection{The algorithm}\label{mhe-alg}

In order to specify the model, we proceed in the following way:

\begin{enumerate}
\item The model is defined by two functions on the
past null cone: the angular diameter distance, $D_A(z)$, and the mass density in
redshift space, $m(z)n(z)$.  We assume that these functions are the same as in
the $\Lambda$CDM model:
\begin{eqnarray}
&& D_A(z) = \frac{1}{1 + z} \frac{c}{H_0} \int_0^z \frac{
 {\rm d} z'}{\sqrt{\Omega_m (1 + z')^3 + 1-\Omega_{m}}}, \label{dalcdm} \\
&&  m(z) n(z) =  \Omega_m \frac{3H_0^2}{8 \pi G} (1+z)^3 D_A^2  \frac{{\rm d}
r}{{\rm d} z},
\end{eqnarray}
where $\Omega_m = (8 \pi G)/(3H_0^2) \rho_0$.

\item Using the MHE algorithm we find $r(z)$ by solving (\ref{eq:rz_int}).

\item We numerically
invert this relation to find $z(r)$ and solve (\ref{mheM}) to find $M(r)$.

\item The function $E(r)$ is found by solving (\ref{mheE}).

\item Once $E$ and $M$ are known, we find $\eta$ and then $t_B$,
by solving the appropriate relations (\ref{2.5})--(\ref{2.8}).

\item Since in (\ref{mheE}) the term $(1-2M/R)/(\dril R z)$ becomes 0/0 at the
apparent horizon, the computer produces inaccurate results in the vicinity. To
overcome this we apply the procedure described in Sec. \ref{mhe-ahp}.

\end{enumerate}

\subsubsection{The results}\label{mhe-res}

The algorithm described in the previous section allows us to find an L--T model
from a given $(D_A(z), m(z)n(z))$ set of data.

\begin{figure}
\begin{center}
\includegraphics[scale=0.65]{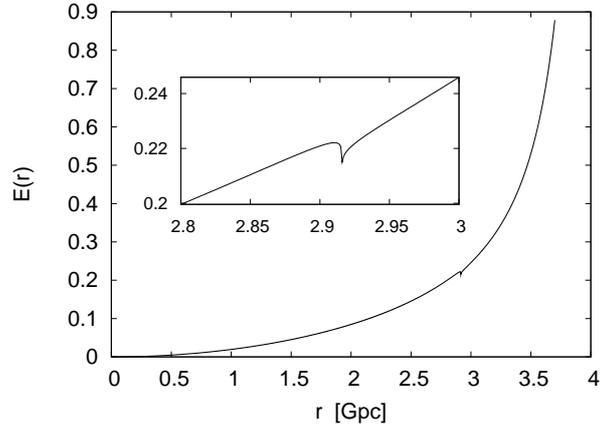}
\caption{The function $E(r)$ of the L--T model defined by the $(D_A, m(z) n(z))$
set of data (see Sec. \ref{mhe-alg1} for details). There is a problem with
numerical integration around $r=2.9$ Gpc. The problem comes from (\ref{mheE})
where the term $(1-2M/R)/(\dril R z)$ becomes 0/0 at the apparent horizon.}
 \label{s1_E}
\end{center}
\end{figure}

\begin{figure}
\begin{center}
\includegraphics[scale=0.65]{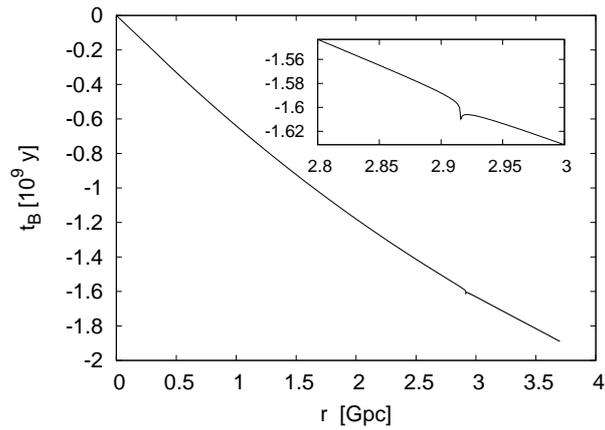}
\caption{The function $t_B(r)$ of the L--T model defined by the $(D_A, m(z)
n(z))$ set of data (see Sec. \ref{mhe-alg1} for details). There is a cusp around
$r=2.9$ Gpc. The cusp follows from an unstable behaviour of $E(r)$ around
$r=2.9$ Gpc.}
 \label{s1_t}
\end{center}
\end{figure}

\begin{figure}
\begin{center}
\includegraphics[scale=0.65]{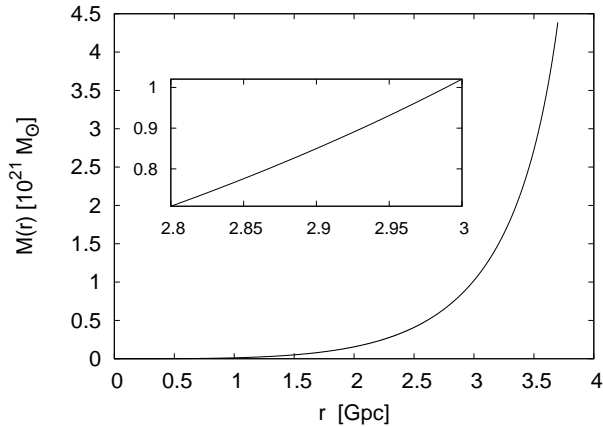}
\caption{The function $M(r)$ of the L--T model defined by the $(D_A, m(z) n(z))$
set of data (see Sec. \ref{mhe-alg1} for details). In this case there is no
problem around $r=2.9$ Gpc.}
 \label{s1_M}
\end{center}
\end{figure}

The free functions of the L -- T model, $E$, $t_B$, and $M$ are shown in Figs.
\ref{s1_E}, \ref{s1_t}, and \ref{s1_M} respectively.

As can be seen, there is a problem with numerical integration for $E$ and $t_B$
around $r=2.9$ Gpc. The problem is related to (\ref{mheE}) where the term
$(1-2M/R)/(\dril R z)$ becomes 0/0 at the apparent horizon. Because of this, the
computer produces inaccurate results in the vicinity. One solution to this
problem was proposed by Lu \& Hellaby (2007) who performed series expansions of
$R(z)$, $n(z)$, $\dril r z$, $M(z)$ and $E(z)$ around the apparent horizon.
However, this method leads either to jumps in one of these functions, say
$E(z)$, or to lower accuracy of the algorithm (Lu \& Hellaby 2007). Therefore,
we propose a different, much simpler approach. Namely, we fit polynomials to
$E(r)$ and $M(r)$ and then we recalculate the area distance and the
redshift-space mass-density as functions of redshift to check the accuracy of
our approximations. As we will see, this method leads to results that from the
observational point of view (every observation is accompanied with an error) are
indistinguishable from those of the $\Lambda$CDM model.

\subsubsection{Dealing with the apparent horizon}\label{mhe-ahp}

To overcome the numerical problem of indeterminacy of $E(r)$ at the apparent
horizon, we fit polynomials to the obtained $M(r)$ and $E(r)$. The most obvious
choice would be polynomials in the variable $r$. However, in numerical
experiments we noticed that much better results are obtained when we approximate
$M$ and $E$ by polynomials in $R(t_0, r)$, where $t_0$ is the present instant.
The explicit forms of the fitted functions are
 \begin{equation}\label{approxM}
M(R(t_0, r)) = \sum_{n=3}^{n=8} M_n {\ell}^n,
 \end{equation}
where ${\ell} = R(t_0,r)/1$Gpc, $(M_3,M_4,M_5,M_6,M_7,M_8)
       = (8.142244  \times 10^{-3}$ kpc, $0$ kpc,
      $1.32458 \times 10^{-4}$ kpc,
      $-3.79   \times 10^{-5}$ kpc,
      $3.23834  \times 10^{-6}$ kpc,
      $-9.8233 \times 10^{-8}$, kpc), and
 \begin{equation}\label{approxE}
E(R(t_0, r)) = \sum_{n=2}^{n=6} E_n {\ell}^n,
 \end{equation}
where $(E_2, E_3, E_4, E_5, E_6)
     = (1.9475  \times 10^{-2}$,
      $4.28698  \times 10^{-3}$,
      $6.50383  \times 10^{-4}$,
      $-3.66095  \times 10^{-5}$,
      $8.78679  \times 10^{-7}).$

\begin{figure}
\begin{center}
\includegraphics[scale=0.65]{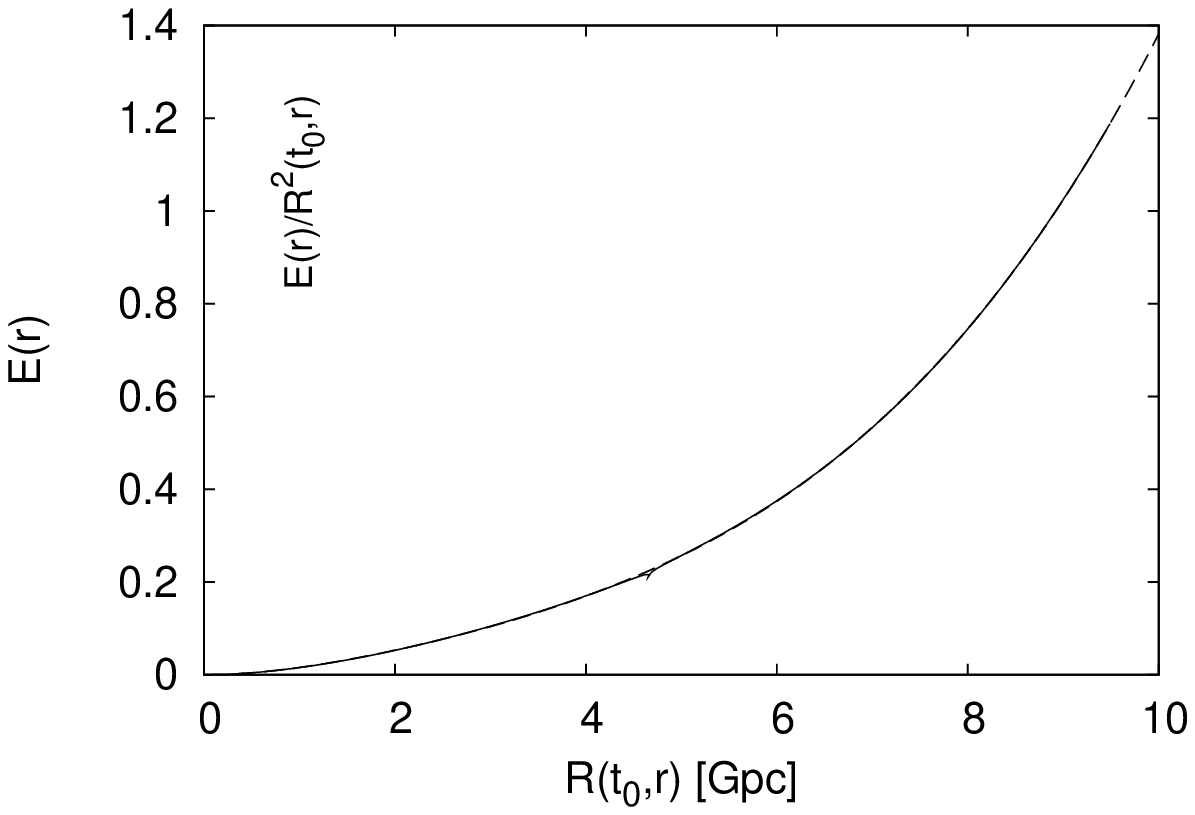}
\caption{The function $E$ as a function of the current areal radius for the
$(D_A, m(z) n(z))$ set of data -- the comparison of results obtained for $E$ as
given by (\ref{mheE}) (solid line) and for the approximation to $E$ as given by
(\ref{approxE}) (dashed line). See Sec.~\ref{mhe-ahp} for details. The inset
presents $E/R^2$, the quantity that is constant in the FLRW limit.}
 \label{s1_Ec}
\end{center}
\end{figure}

\begin{figure}
\begin{center}
\includegraphics[scale=0.65]{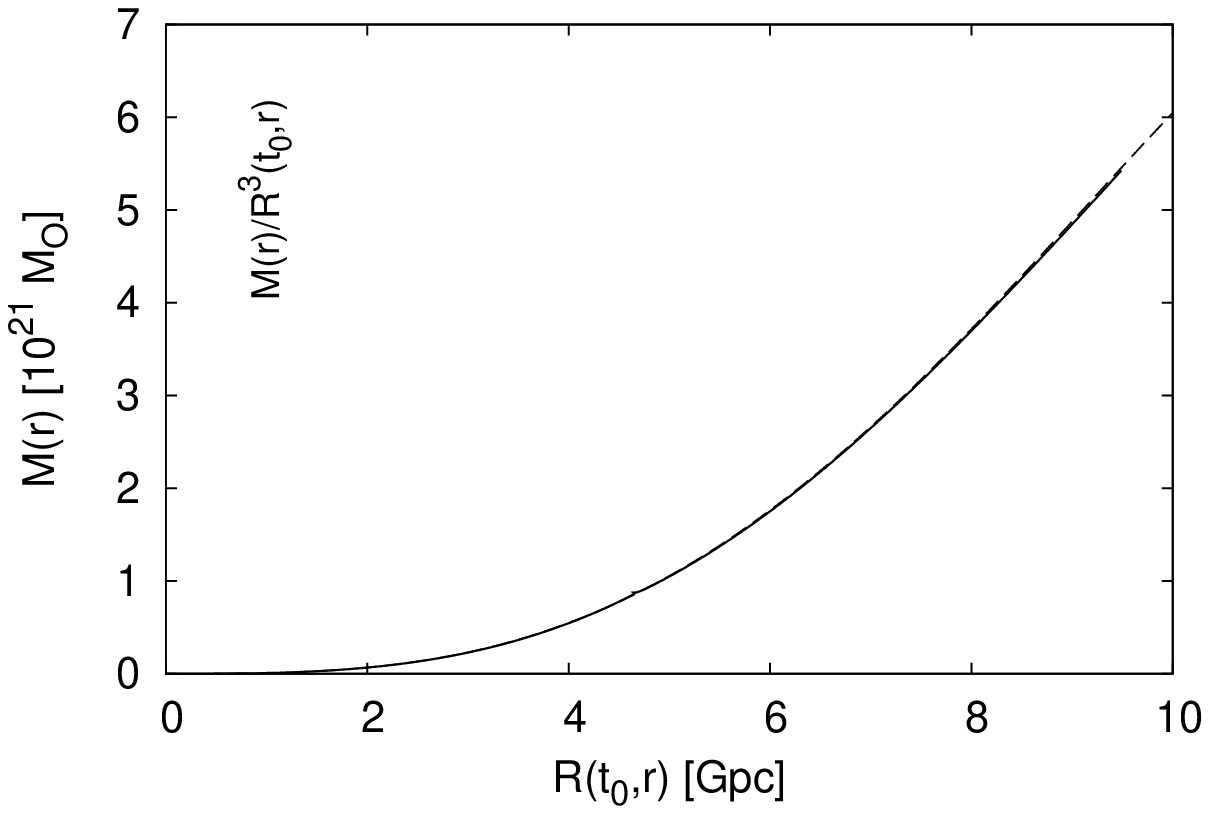}
\caption{The function $M$ as a function of the current areal radius for the
$(D_A, m(z) n(z))$ set of data -- the comparison of results obtained for $M$ as
given by (\ref{mheM}) (solid line) and for the approximation to $M$ as given by
(\ref{approxM}) (dashed line). See Sec.~\ref{mhe-ahp} for details. The inset
presents $M/R^3$, the quantity that is constant in the FLRW limit.}
 \label{s1_Mc}
\end{center}
\end{figure}

The profiles of these functions together with those numerically derived are
presented in Figs. \ref{s1_Ec} and \ref{s1_Mc}. We then use these functions as
initial conditions and solve the null geodesic equations. First, we invert
(\ref{2.11}) and (\ref{2.12}) to get the equations for ${\rm d} t/{\rm d} z$ and
${\rm d} r/{\rm d} z$, to derive the pair $(t,r)$ for a given redshift;
simultaneously we solve (\ref{2.2}) to get $\widehat{R}$, $\widehat{R,_r}$, and
$\widehat{R,_{tr}}$.

\begin{figure}
\begin{center}
\includegraphics[scale=0.65]{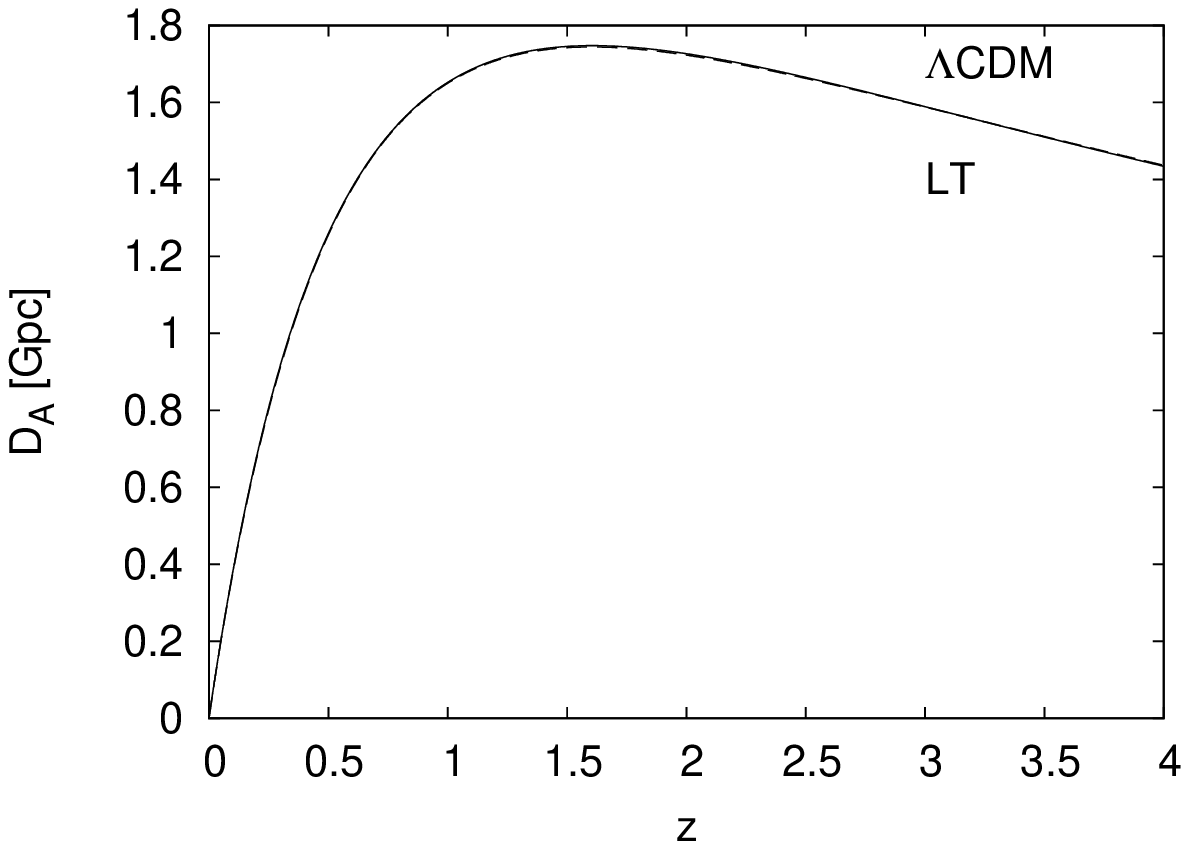}
\caption{The angular diameter distance as a function of redshift; comparison of
the results for the $\Lambda$CDM and L--T models. The inset presents the
estimations of $D_A$ based on the type-Ia supernova measurements taken from the
Union data set (Kowalski {\it et al.} 2008). See Sec.~\ref{mhe-ahp}}.
 \label{s1_da}
\end{center}
\end{figure}

\begin{figure}
\begin{center}
\includegraphics[scale=0.65]{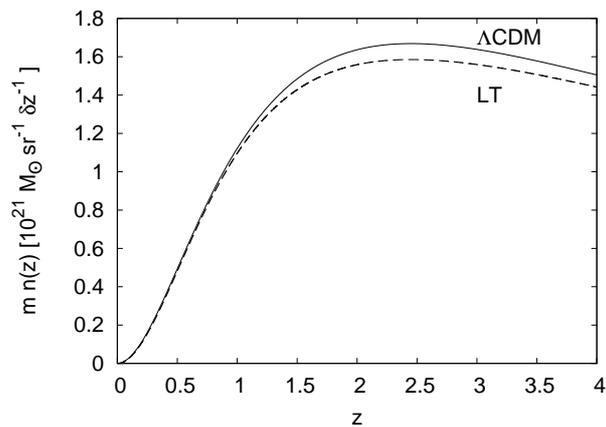}
\caption{The redshift-space mass-density as a function of redshift.
The difference between the LT and $\Lambda$CDM models is less than $6\%$.
See Sec.~\ref{mhe-ahp}. } \label{s1_nc}
\end{center}
\end{figure}

\begin{figure}
\begin{center}
\includegraphics[scale=0.65]{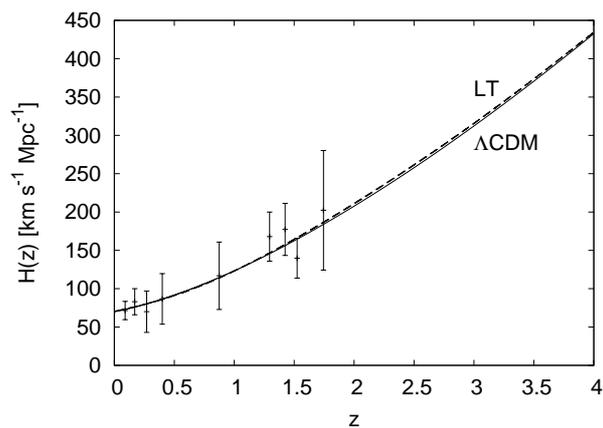}
\caption{The function $H(z)$. See Sec.~\ref{mhe-ahp} for details. For comparison
the measurements of $H(z)$ (Simon {\em et al.} 2005) are also shown. The
difference between the LT and $\Lambda$CDM models is less than $5\%$.}
\label{s1_hz}
\end{center}
\end{figure}

\begin{figure}
\begin{center}
\includegraphics[scale=0.65]{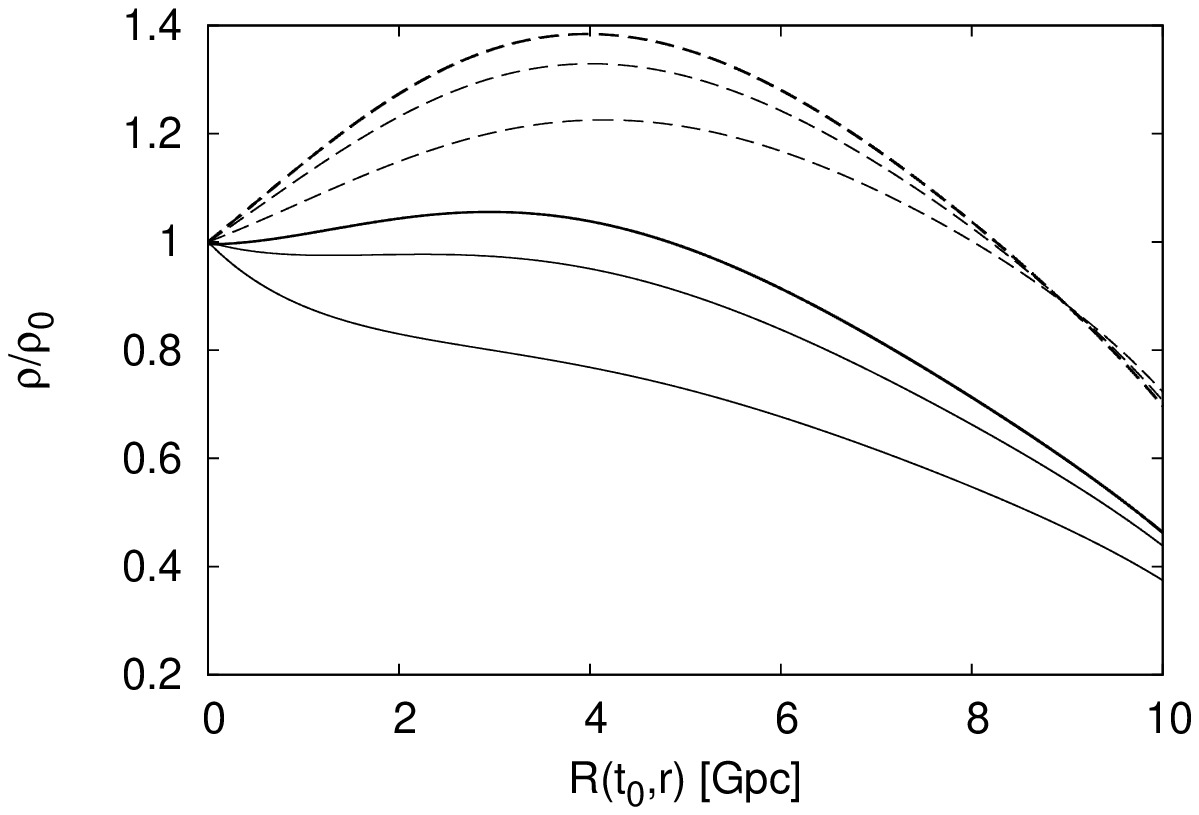}
\caption{The ratio $\rho / \rho_0$ of density to density at the origin.
Solid lines present density profiles at constant $t$ (from top to bottom: $t=
now$, $t=now -2 \times 10^9$ y, $t=now -5 \times 10^9$ y). Dashed lines present
density profiles at the hypersurface of constant age of the Universe (from top
to bottom: $\tau = now$, $\tau=now -2 \times 10^9$ y, $\tau=now -5 \times
10^9$ y). For each curve, the $\rho_0$ is taken at the value of $t$ or,
respectively, $\tau$ that identifies the curve. Note: all the graphs, for both foliations, use a comoving radial coordinate, $r:= R(t_0, r)$, which means that points with the same horizontal coordinate in the graph correspond to the same matter particle at all times. Only for the uppermost solid graph is $R(t_0, r)$ equal to the actual area-distance from the centre.}
 \label{s1_rho}
\end{center}
\end{figure}

The results are shown in Figs. \ref{s1_da}--\ref{s1_nc}. The angular diameter
distance is recovered very accurately, while the redshift-space mass-density
less so, but still up to $z=4$ it does not differ by more than $6\%$ from the
redshift-space mass-density in the $\Lambda$CDM model --- which is far less than
the expected observational uncertainty. In addition we calculate the prediction
for $H(z)$, and we compare it to the estimations of the expansion rate by Simon
{\em et al.} (2005). Since these are based on the observed age of the oldest
stars, and $H(z)$ follows from $\dril t z$, thus, as seen from (\ref{2.12}),
$H(z) = {\widehat{R,_{tr}}}/{\widehat{R,_r}}$. The results are presented in Fig.
\ref{s1_hz}. As seen, the L--T model does not deviate from the $\Lambda$CDM
model by more than $5\%$. These differences in $m(z) n(z)$ and $H(z)$ are caused
by two factors: a) the equations (\ref{approxM}) and (\ref{approxE}) are just
approximations, b) numerical errors in the vicinity of the apparent horizon bias
the solution of (\ref{mheE}) for $r\gtrsim r_{AH}$ (where $r_{AH}$ is the
position of the apparent horizon). In principle, however, it is possible to
construct the $\Lambda=0$ L--T model that matches the $\Lambda$CDM model.
Finally, as seen from Fig. \ref{s1_rho}, the current density profile does not
exhibit a giant void shape. Instead, it suggests that the universe smoothed out
around us with respect to directions is overdense in our vicinity up to
Gpc-scales. As a consequence of our numerical procedure the value of density at
the centre at the present time $t_0$ is the same  as the present density in the
$\Lambda$CDM model.

\subsection{The model defined by $D_A(z)$ and $H(z)$}\label{DHmodel}

\subsubsection{The algorithm}\label{dlhz-alg}

The algorithm used to find the L--T model consists of the following steps:

\begin{enumerate}
\item The model is defined by two functions on the
past null cone: the angular diameter distance $D_A(z)$ and the Hubble function
$H(z)$. We assume that these functions are the same as in the $\Lambda$CDM
model -- $D_A(z)$ is given by (\ref{dalcdm}) and
\begin{equation}
H(z) = H_0 \sqrt{{\Omega_m (1 + z)^3 + 1-\Omega_{m}}}.\label{hzlcdm}
\end{equation}
\item We choose $r$ so that (\ref{corchoice}) is satisfied
on the past light cone of the present-day observer. Then using $H(z) =
{\widehat{R,_{tr}}}/{\widehat{R,_r}}$, (\ref{2.11}) becomes
 \begin{equation}
\frac{{\rm d} r}{{\rm d} z} = \frac{1}{1+z} \frac{c}{H(z)},
\end{equation}
which for (\ref{hzlcdm}) can be integrated to
\begin{eqnarray}
r = && \frac{2c}{3H_0} \frac{1}{\sqrt{1 - \Omega_m}} \left[
{\rm arsinh} \left( \sqrt{ \frac{1-\Omega_m}{\Omega_m}} \right)
\right. \nonumber \\
&& \left. -
{\rm arsinh} \left( \sqrt{ \frac{1-\Omega_m}{\Omega_m (1+z)^3}} \right) \right].
\end{eqnarray}

\item Using (\ref{nullhz}) we find $\rho(z)$ and solve (\ref{eq:rho_mn}) for
$n(z)$.

\item We solve (\ref{mheM}) to find $M(r)$.

\item The function $E(r)$ is found by solving (\ref{mheE}).

\item Once $E$ and $M$ are known we find $\eta$ and then $t_B$
by solving the appropriate relations (\ref{2.5})--(\ref{2.8}).

\item As before, because of the 0/0 term in (\ref{mheE}) at the apparent
horizon, we employ the procedure described in Sec. \ref{dlhz-ahp}.
\end{enumerate}

\subsubsection{The results}\label{dlhz-res}

\begin{figure}
\begin{center}
\includegraphics[scale=0.65]{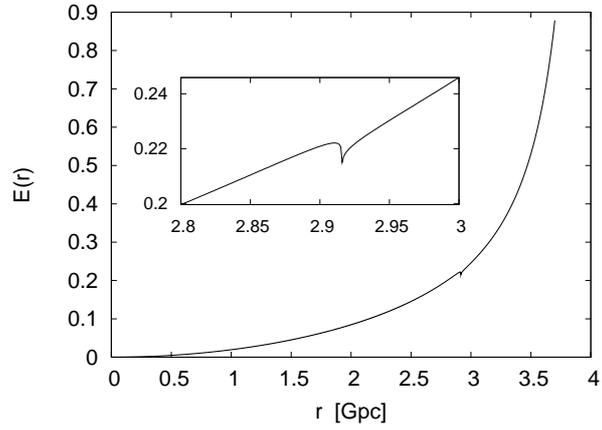}
\caption{The function $E(r)$ of the L--T model defined by the $(D_A, H(z))$ set
of data (see Sec. \ref{dlhz-alg} for details). Around $r=2.9$ Gpc there is a
problem with the numerical algorithm.}
 \label{s2_E}
\end{center}
\end{figure}

\begin{figure}
\begin{center}
\includegraphics[scale=0.65]{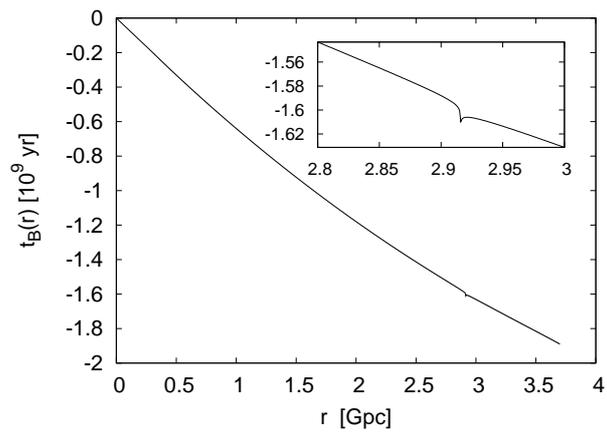}
\caption{The function $t_B(r)$ of the L--T model defined by the $(D_A, H(z))$
set of data (see Sec. \ref{dlhz-alg} for details).}
 \label{s2_t}
\end{center}
\end{figure}

\begin{figure}
\begin{center}
\includegraphics[scale=0.65]{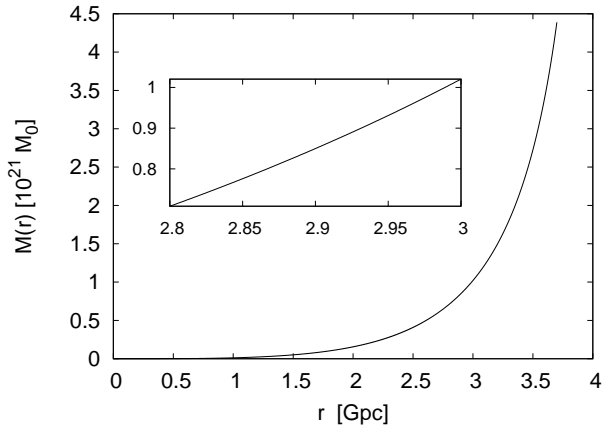}
\caption{The function $M(r)$ of the L--T model defined by the $(D_A, H(z))$ set
of data (see Sec. \ref{dlhz-alg} for details). In this case there is no problem
around $r=2.9$ Gpc.}
 \label{s2_M}
\end{center}
\end{figure}

The results for $E$, $t_B$ and $M$ are given in Figs.~\ref{s2_E} -- \ref{s2_M}.
As in Sec.~\ref{mhe-res}, the functions $E(r)$ and $t_B(r)$ evaluated by this
algorithm behave unnaturally close to the apparent horizon, see Figs.~\ref{s2_E}
and \ref{s2_t}. As before, this is caused by the fact that in (\ref{mheE}) one
has to deal with 0/0. We overcome this problem by once again fitting polynomials
to these functions.

\subsubsection{Dealing with the apparent horizon}\label{dlhz-ahp}

The explicit forms of the fitted functions are:
 \begin{equation}\label{Mapp_dlhz-ahp}
M(R(t_0,r)) = \sum_{n=3}^{n=7} M_n {\ell}^n,
 \end{equation}
where $(M_3,M_4,M_5,M_6,M_7) =$ $(8.2 \times 10^{3}$kpc, $-1.1948$ kpc, $1.07521
\times 10^{2}$ kpc, $-24.1385$ kpc, $1.15743$  kpc), and
 \begin{equation}\label{Eapp_dlhz-ahp}
E(R(t_0,r)) = \sum_{n=2}^{n=6} E_n {\ell}^n,
 \end{equation}
where $(E_2,E_3,E_4,E_5,E_6) =(1.7324 \times 10^{-2}, -2.5725 \times 10^{-3},$
$1.14925 \times 10^{-4}, 2.87776 \times 10^{-5}, -1.90389 \times 10^{-6}).$

\begin{figure}
\begin{center}
\includegraphics[scale=0.65]{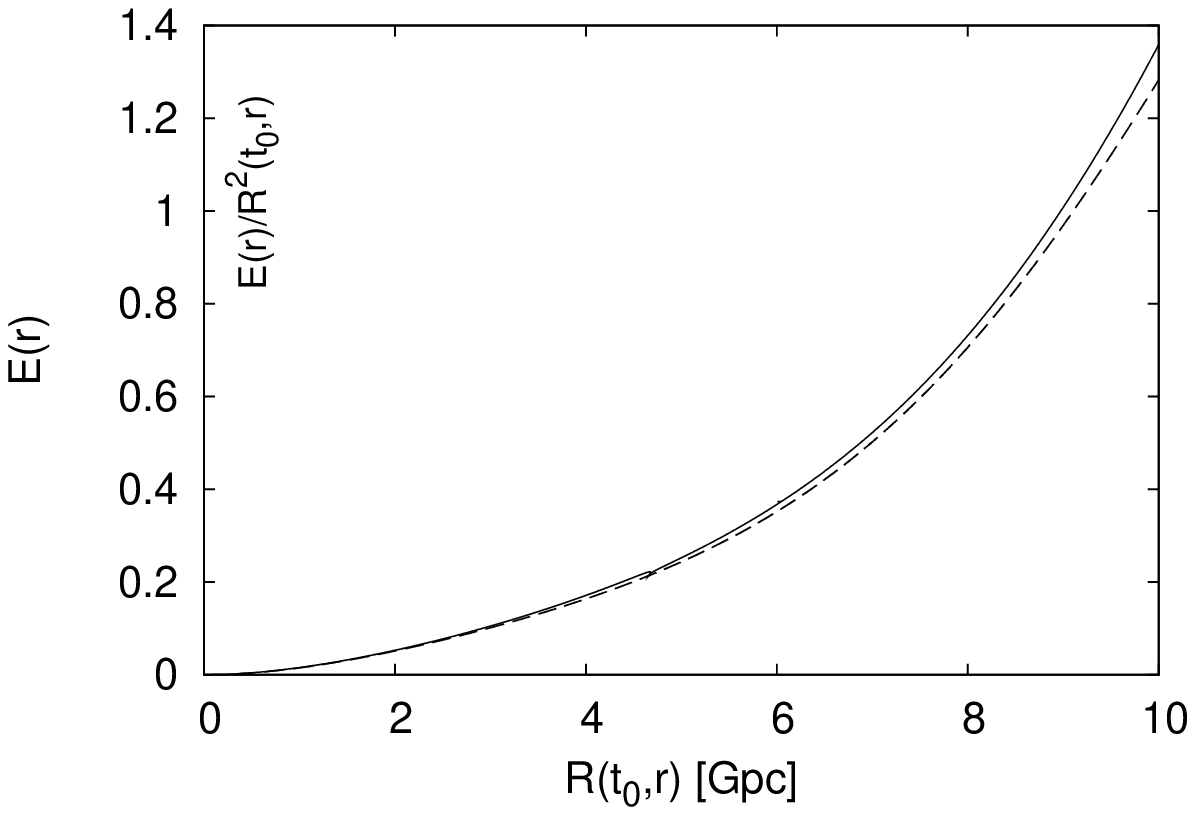}
\caption{The function $E$ as a function of the current areal radius for the
$(D_A, H(z))$ set of data -- the comparison of results obtained for $E$ as given
by (\ref{mheE}) (solid line) and for the approximation to $E$ as given by
(\ref{Eapp_dlhz-ahp}) (dashed line). See Sec.~\ref{dlhz-ahp} for details.
The inset presents $E/R^2$, the quantity that is constant in the FLRW limit.}
 \label{s2_Ec}
\end{center}
\end{figure}

\begin{figure}
\begin{center}
\includegraphics[scale=0.65]{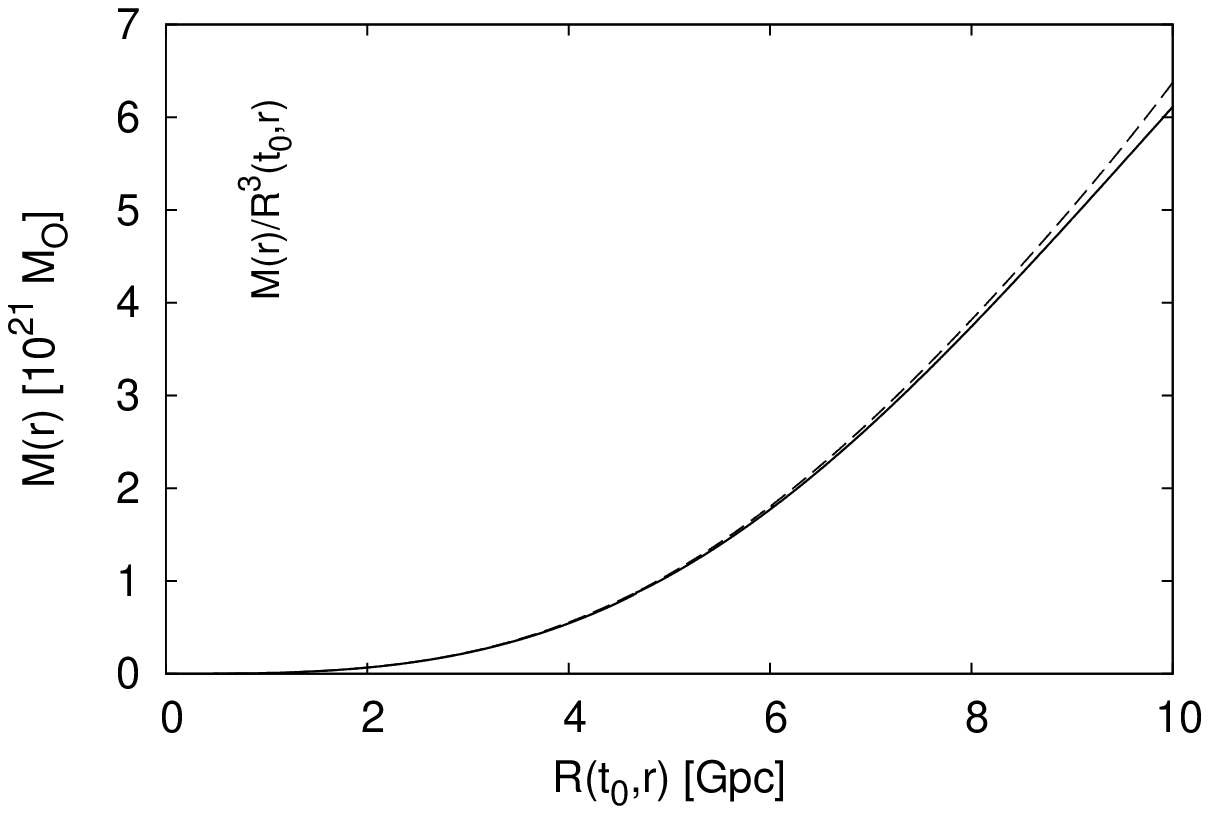}
\caption{The function $M$ as a function of the current areal radius for the
$(D_A, H(z))$ set of data -- the comparison of results obtained for $M$ as given
by (\ref{mheM}) (solid line) and for the approximation to $M$ as given by
(\ref{Mapp_dlhz-ahp}) (dashed line). See Sec.~\ref{dlhz-ahp} for details.
The inset presents $M/R^3$, the quantity that is constant in the FLRW limit.}
 \label{s2_Mc}
\end{center}
\end{figure}

\begin{figure}
\begin{center}
\includegraphics[scale=0.65]{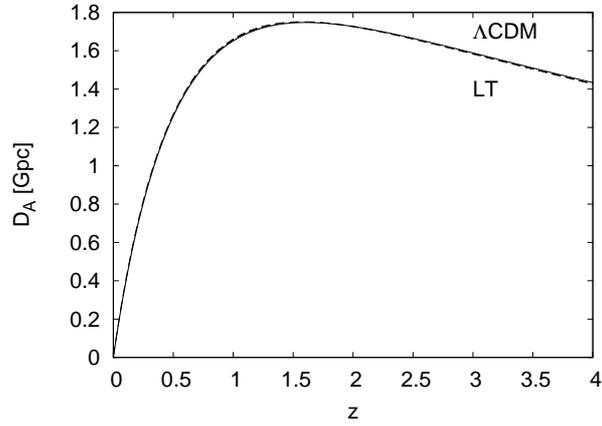}
\caption{The angular diameter distance as a function of $z$. See
Sec.~\ref{dlhz-ahp} for details.}
 \label{s2_da}
\end{center}
\end{figure}

\begin{figure}
\begin{center}
\includegraphics[scale=0.65]{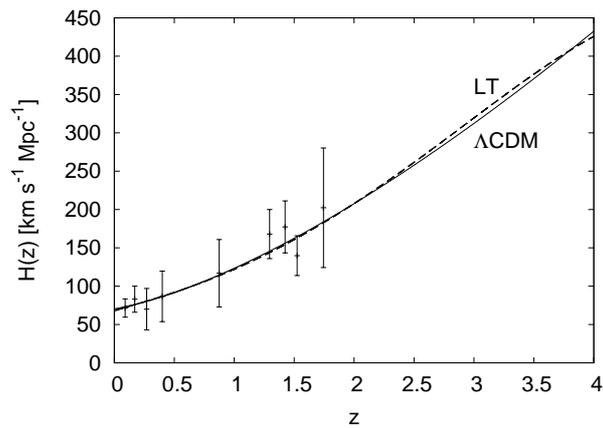}
\caption{The function $H(z)$. See Sec.~\ref{dlhz-ahp} for details. For
comparison the measurements of $H(z)$ (Simon {\em et al.} 2005) are also shown.}
 \label{s2_hz}
\end{center}
\end{figure}

\begin{figure}
\begin{center}
\includegraphics[scale=0.65]{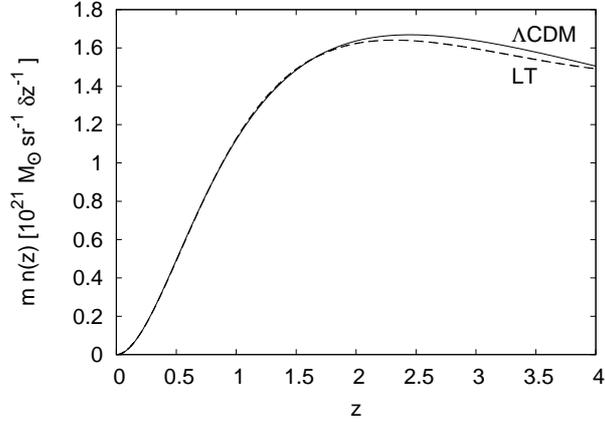}
\caption{ The redshift-space mass-density as a function of redshift for the LT
model considered in Sec.~\ref{dlhz-ahp}, and in the $\Lambda$CDM model. The
difference between the two models is less than 2.5\%. Compare Fig, \ref{gnc};
the consistency achieved here is much better than in a giant void model.}
\label{s2_nc}
\end{center}
\end{figure}

\begin{figure}
\begin{center}
\includegraphics[scale=0.65]{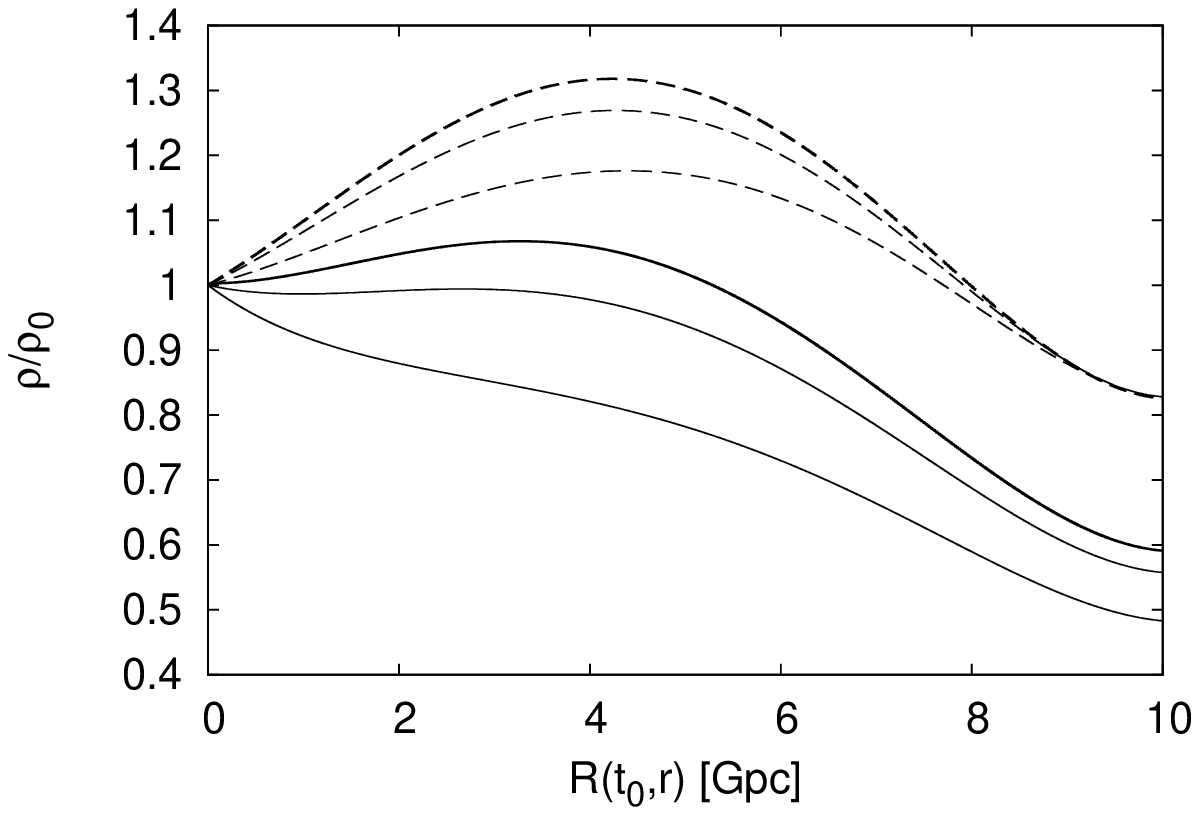}
\caption{The ratio $\rho / \rho_0$ of density to density at the origin.
Solid lines present density profiles at constant $t$ (from top to bottom: $t=
now$, $t=now -2 \times 10^9$ y, $t=now -5 \times 10^9$ y). Dashed lines present
density profiles at the hypersurface of constant age of the Universe (from top
to bottom: $\tau = now$, $\tau=now -2 \times 10^9$ y, $\tau=now -5 \times
10^9 y$). For each curve, the $\rho_0$ is taken at the value of $t$ or,
respectively, $\tau$ that identifies the curve. The remark about the horizontal coordinate in Fig. \ref{s1_rho} applies also here.}
 \label{s2_rho}
\end{center}
\end{figure}

The profiles of these functions together with the numerically derived ones are
shown in Figs.~\ref{s2_Ec} and \ref{s2_Mc}. We then use these functions as
initial conditions and solve the null geodesic equations. The results are
presented in Figs.~\ref{s2_da} and \ref{s2_hz}. Both the angular diameter
distance and the expansion rate as functions of the redshift are recovered very
accurately. From the observational perspective these two models are
indistinguishable. In addition we present the $m(z)n(z)$ plot. As seen from Fig.
\ref{s2_nc}, it also gives quite an accurate fit, with a deviation from the
$\Lambda$CDM model being less than $2.5\%$. Finally, as seen from Fig.
\ref{s2_rho}, the current density profile has a similar shape as in Sec.
\ref{mhe-ahp}, and this is not a giant void.

\section{Discussion and conclusion} \label{disconc}

Contrary to what is commonly claimed, L--T models with a giant void do not
reproduce the main features of the $\Lambda$CDM model. These types of models
just fit cosmological observations, with a priori constraints imposed on the
L--T models. Indeed, we have found the L--T models that mimic some of the
observational features of the $\Lambda$CDM model and exhibit no giant void, but
rather a giant hump.

It is clear from the energy density profiles shown in Figs. \ref{s1_rho} and \ref{s2_rho} that the L-T models we have reconstructed have a large overdensity, when viewed over large enough scales. However, as we have made clear throughout, these profiles are the result of reproducing observables that match $\Lambda$CDM predictions, and not from fitting to any real data. Luminosity distances from supernovae observations, and number counts from galaxy surveys do not currently extend much beyond $z \sim 1.5$ and $z \sim 0.5$,
respectively. As such, using real observables, it would only currently be possible to perform a reconstruction of a limited part of the full structure we have found here, and even then only up to the degree allowed by the errors associated with these quantities. If one were to attempt such a reconstruction, it appears from Figs. \ref{s1_rho} and \ref{s2_rho} that one may, in fact, reconstruct a local energy density profile that is an increasing function of $r$, and not decreasing, due to the limited extent of these observations in $z$ (this is particularly true in the equal age foliation). Our interpretation of a giant hump should therefore be understood as corresponding to an extrapolation of observable quantities under the expectation that they will follow $\Lambda$CDM, rather than being directly implied by any currently known observations themselves.

Recently, some astrophysicists have begun to take seriously the cosmological
implications of the L--T model. This model, although still simple,\footnote{From
the computational point of view, the difference between the Friedmann and L--T
models is quite trivial. The equation (\ref{2.2}) that governs the evolution of
the L--T model is exactly the same as in the Friedmann model; it is still an
{\em ordinary} differential equation in the time-variable $t$. The only
difference is that the function $R(t)$ obeying (\ref{2.2}) depends on one more
variable, the radial coordinate $r$. This $r$ enters only as a parameter, and
then automatically all the integration `constants' that appear while solving
(\ref{2.2}) are no longer constant, but are functions of $r$. Sophistication
comes at the level of interpreting the solutions -- however, this is no longer
mathematics, but astrophysics.} is quite powerful and exhibits some features of
general relativistic dynamics, like arbitrary functions in the initial data.

As we said earlier in this paper, the belief that an L--T model fitted to
supernova Ia observations necessarily implies the existence of a giant void with
us at the centre was created by arbitrarily limiting the generality of the model. With its free functions fitted to $\Lambda$CDM features rather than to expectations, the giant void does not necessarily follow. Rather, one alternative is that the graph of the density smoothed out over angles around us has the shape of a shallow and wide valley on top of a giant hump.

This giant hump may be a feature of the particular L--T model that we ended up
with. Variations in the $\Lambda$CDM parameters would modify the details, and
the observational constraints we considered are not very tight. Future
calculations with other constraints may favor a still different profile at $t =$
now. Hence we do not wish our paper to become a starting point of a new paradigm
in observational cosmology, aimed at detecting the hump. Before this happens, we
must decide {\em at the theoretical level} whether the hump is a necessary
implication of L--T models properly fitted to other observations.

\begin{figure}
\begin{center}
\includegraphics[scale=0.65]{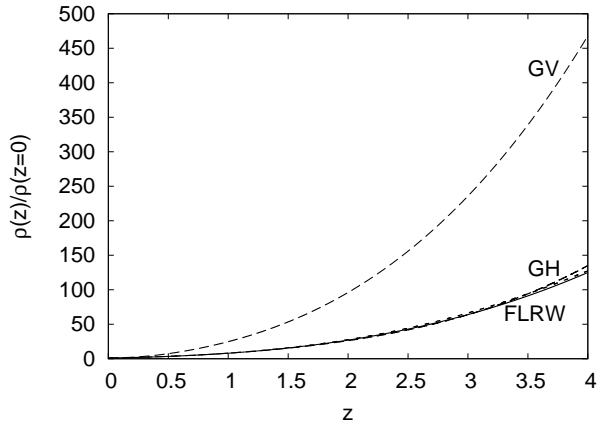}
\caption{The density distribution as a function of redshift for the giant
void (GV) L--T model, for the models of Sec.~\ref{mhe-ahp} and \ref{dlhz-ahp}
(GH) and for the FLRW model, where $\rho(z) = \rho_0(1+z)^3$ (dashed line). The
parameters of the GV model are: radius 2.96 Gpc and density contrast 4.05. The
same model is presented in Fig. 1; for details of how the model is specified and
null gedesics solved see Bolejko \& Wyithe 2009.}
 \label{drs}
\end{center}
\end{figure}

It must be stressed that this hump {\em is not directly observable}. It exists
in the space $t =$ now, of events simultaneous with our present instant in the
cosmological synchronisation, i.e. it is in a space-like relation to us. This is
also the case of the giant void (see, e.g., Fig. 4 of Alnes {\it et al.} 2006;
Fig. 1 of Garc\'ia-Bellido and Haugb\o lle 2008a; Figs. 4 and 6 of Yoo {\it et
al.} 2008). However, an observational test of the giant void is easier to
complete. The reason is that the models considered in this paper have a
redshift-space mass-density almost the same as in the $\Lambda$CDM model (see
Figs.~\ref{s1_nc} and \ref{s2_nc}) and as can be seen from Fig.~\ref{drs}, their
$\rho(z)$ scales with the redshift in almost the same manner as in the FLRW
model, i.e. $\sim (1+z)^3$. This is not the case of the giant void\footnote{The
giant void used here is Bolejko \& Wyithe (2009)'s model with radius of 2.96 Gpc
and density contrast of 4.05. The redshift-space mass-density for this model is
presented in Fig. 1.} which does not reproduce these features on the past light
cone (see Figs.~\ref{gnc} and \ref{drs}). Thus, unlike the giant void, the giant
hump is not observable in $\rho(z)$ or in the number count data.

What is the cause of this difference between the density distribution on our
past light cone and in the $t =$ now space? It is the oft-forgotten basic
feature of the L--T model (and in fact of all inhomogeneous models, also those
not yet known explicitly as solutions of Einstein's equations): on any initial
data hypersurface, whether it is a light cone or a $t =$ constant space, {\em
the density and velocity distributions are two algebraically independent
functions of position}. Thus the density on a later hypersurface may be quite
different, since it depends on both initial functions. Whatever initial density
distribution we observe can be completely transformed by the velocity
distribution. For example, as predicted by Mustapha and Hellaby (2001) and
explicitly demonstrated by Krasi\'nski and Hellaby (2004), any initial
condensation can evolve into a void and vice versa. In FLRW models, there are no
physical functions of position, and all worldlines evolve together. Thus, while
dealing with an L--T (or any inhomogeneous) model, one must be cautious when applying the Robertson--Walker-inspired prejudices and expectations.

The use of oversimplified L--T models can create another false idea and false
expectation. The false idea is that there is an opposition between the
$\Lambda$CDM model, belonging to the FLRW class, and the L--T model or in general, inhomogeneous models: either one or the other could be `correct', but not both. This putative opposition can then give rise to the expectation that more, and more detailed, observations will be able to tell us which one to reject. In truth, there is no opposition. The inhomogeneous models, like for example the L--T
model with its two arbitrary functions of one variable are huge, compared to
FLRW, {\em families} of models that include the Friedmann models as a very
simple subcase. The fact, demonstrated in several papers already (see
C\'el\'erier 2007, for a review), that even a $\Lambda = 0$ L--T model can
imitate $\Lambda \neq 0$ in an FLRW model, additionally attests to the
flexibility and power of the L--T model. Thus, if the Friedmann models,
$\Lambda$CDM among them, are considered good enough for cosmology, then the L--T
models can only be better: they constitute an {\em exact perturbation} of the
Friedmann background, and can reproduce the latter as a limit with an arbitrary
precision. While future observations, for example the kSZ effect
(Garcia-Bellido \& Haugb\o lle 2008b) or the growth of linear structure (Clarkson {\it et al.} 2009)) will provide a sufficient insight to test particular configurations (like for example a giant void model), we will never be able to reject inhomogeneous models. After all, the Universe as it is, is inhomogeneous. Nowadays we use homogeneous models just for simplicity, and although they have worked well so far, in future they will certainly be replaced by more sophisticated models, either by exact solutions, or what is more probable in light of increasing computation power of computers, by numerical simulations.

When considering models that go beyond the FLRW approximation, one may ask either `what limitations on the arbitrary functions in the models do our observations impose', or `which model best describes a given situation: a homogeneous FLRW model or an inhomogeneous one?' The latter of these questions has often been asked in the context of comparing L-T models without $\Lambda$ to FLRW models with $\Lambda$, and is of much interest for understanding the necessity of introducing $\Lambda$ into the observer's standard cosmological model. Such hypothesis testing questions are often posed in cosmology, but are difficult to address in the current context as they require artificially limiting the generality of the models in question (in order to have a finite number of parameters, so that the test can be performed). Given the lack of motivation for exactly how to perform such a limitation, one is then left in the undesirable circumstance of (potentially) dismissing particular L-T models, while being left with an infinite number of remaining L-T models to evaluate. We therefore prefer to consider the former question. In order to reasonably answer this for the L--T model,
a general framework for interpreting observations in the L--T geometrical
background should be created (and in the future it should be transformed into a
framework for interpreting the observations in a still more general, or {\em the
most general} geometrical background). Such a program is still in its infancy,
but is being actually developed by C. Hellaby and coworkers under the name
`Metric of the Cosmos' (Lu and Hellaby, 2007, McClure and Hellaby, 2008, Hellaby
and Alfedeel, 2009).

{\it Acknowledgements:}
MNC wants to acknowledge interesting discussions with Rocky Kolb, Luca Amendola,
Roberto Sussman and Miguel Quartin. We also acknowledge useful comments on
earlier versions of this text by Charles Hellaby.

\end{document}